\documentclass{elsarticle}
\usepackage[T1]{fontenc}
\usepackage[latin9]{inputenc}
\usepackage{amsbsy}
\usepackage{amssymb}
\usepackage{graphicx}

\makeatletter
\usepackage{amsmath}

\makeatother

\begin{document}

\title{Exact results for fixation probability of bithermal evolutionary
graphs.}

\author{Bahram Houchmandzadeh and Marcel Vallade.}

\address{CNRS/Univ. Grenoble 1, LIPhy UMR 5588, Grenoble, F-38041, France.}
\begin{abstract}
One of the most fundamental concepts of evolutionary dynamics is the
``fixation'' probability, \emph{i.e. }the probability that a mutant
spreads through the whole population. Most natural communities are
geographically structured into habitats exchanging individuals among
each other and can be modeled by an evolutionary graph (EG), where
directed links weight the probability for the offspring of one individual
to replace another individual in the community. Very few exact analytical
results are known for EGs. We show here how by using the techniques
of the fixed point of Probability Generating Function, we can uncover
a large class of of graphs, which we term bithermal, for which the
exact fixation probability can be simply computed. \end{abstract}
\maketitle

\section{Introduction.}

Evolutionary dynamics is a stochastic process due to competition between
deterministic selection pressure and stochastic events due to random
sampling from one generation to the other. One of the most fundamental
concepts of evolutionary dynamics is the \emph{fixation probability},
\emph{i.e. }the probability that a mutant spreads and takes over the
whole community(\citet{Patwa2008a}). In the framework of the Moran
model (\citet{Moran1962}) for a well mixed population, where an individual's
offspring can replace any other one in the community, the fixation
probability is

\begin{equation}
\pi_{f}=\frac{1-r^{-m_{0}}}{1-r^{-M}}\label{eq:Moran0}
\end{equation}
where $M$ is the size of the community, $m_{0}$ the original number
of mutants and $r$ the relative fitness of the mutants. A similar
result was reached by Kimura (\citet{Kimura1962}) for the Fisher-Wright
model under the diffusion approximation. 

\begin{figure}
\begin{centering}
\includegraphics[width=0.8\columnwidth]{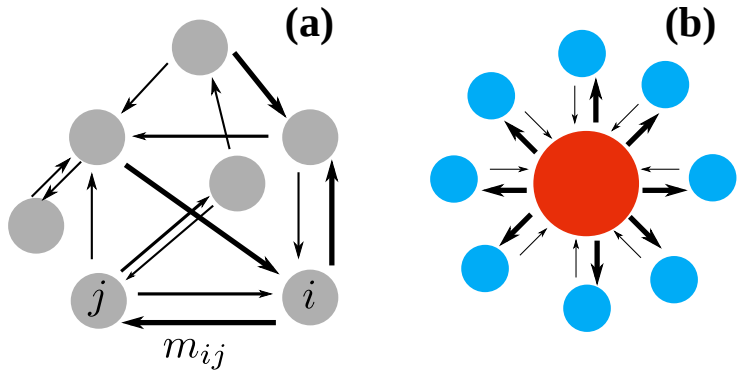}
\par\end{centering}

\caption{Evolutionary Graphs. (a) individuals are spread in space. A node $i$
can send its offspring only to connected nodes $j$ with a probability
$m_{ij}$. A non-structure population can be considered as a fully
connected graph with uniform migration probability $m_{ij}=1/N$.
(b) The star configuration, which can be proved to differ from a non-structured
population for its fixation probability. \label{fig:EGscheme}}

\end{figure}

The idea of well mixed population is however far from realistic. Natural
communities are geographically extended and subdivided into patches
that exchange individuals (figure \ref{fig:EGscheme}a). Maruyama
(\citet{Maruyama1974,Maruyama1974a}) was the first to cast the problem
of evolutionary dynamics into a Moran process on graphs (or islands,
in his terminology) and under harsh approximations, concluded that
the fixation probability does not depend on the population structure.
The first formal proof that Maruyama's results are not always correct
was given by Lieberman \emph{et al}(\citet{Lieberman2005}) who showed
that for a Moran process on a star graph (Figure \ref{fig:EGscheme}b),
the effective fitness of the mutant, in the large population limit,
can be enhanced to $r^{2}$ due to topological effects. In order to
do so, Lieberman \emph{et al. }considered evolutionary graphs (EG)
where nodes contain exactly one individual, whether wild type or mutant,
connected by directed links representing the geographical (or social)
connectivity. 
Lieberman et al. also extended their results to the funnel topology
with $K$ layers where the effective fitness, in the limit of large
population, can be amplified to $r^{K}$, but provided only a sketch
of the proof. Beyond the special cases considered by Lieberman et
al, very few exact analytical results are known. A review of the present
state of known results is given by Shakarian et al(\citet{Shakarian2012}). 

Most of the results of the EG are obtained through Monte Carlo numerical
simulations. These simulations however scale as $2^{M}$ where $M$
is the number of nodes. A new numerical scheme has been proposed (\citet{Barbosa2010})
to accelerate the speed of these simulations, but the computation
of the fixation probability of large graphs is still very time consuming. 

It would therefore be important to know the exact fixation probability
of a large class of graphs that can be used as an approximation of
closely related graphs or as a benchmark for assessing the progress
in numerical simulation schemes. This is the aim of the present article.

We recently proposed a new method (\citet{Houchmandzadeh2011}), based
on the fixed points of the time dependent Probability Generating Function
(fp-PGF) which can efficiently approximate the fixation probability
of large, arbitrary graphs by solving only a system of $M$ second
order algebraic equations. We show in the present article that the
fp-PGF method can also be used to derive \emph{exact} results for
a large class of graphs that we call bithermal, which are an extension
of the isothermal graphs considered by Lieberman et al. 

\begin{figure}
\begin{centering}
\includegraphics[width=0.5\columnwidth]{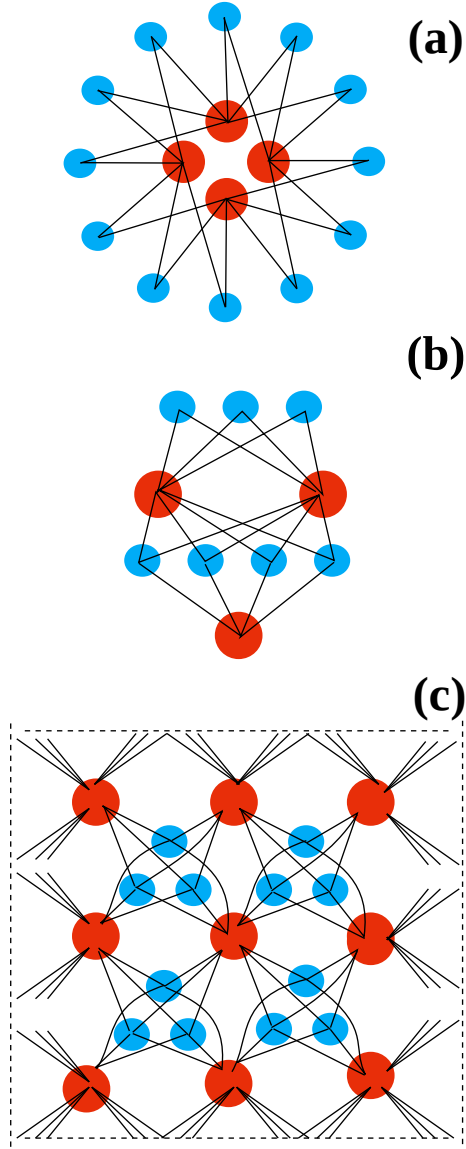}
\par\end{centering}

\caption{Some examples of bithermal graphs; for simplicity, each link represents
two directed links. A bithermal graph is constituted of two kinds
of nodes at respective temperature $T_{A}$ and $T_{B}$, indicated
here by their colors. Each $A$ node is connected to a subset of $B$
nodes and vice versa. (a) a (4,12,6,2) generalized star, consisting
of 4 central nodes and 12 peripheral nodes, each central node connected
to 6 peripheral and each peripheral connected to two central nodes
; (b) an example where the combined effect of the number of links
and their weight makes the graph bithermal ; (c) A symmetric 2 dimensional,
bithermal crystal with periodic boundaries; \label{fig:examples}}

\end{figure}

The \emph{temperature} of a node is related to the imbalance between
the sum of the weights of the incoming and outgoing links. In isothermal
graphs, all nodes are balanced and have the same temperature $T=1$.
\emph{Bithermal} graphs are bipartite, with two kinds of nodes at
either temperature $T_{A}$ or $T_{B}$. The star topology (figure
\ref{fig:EGscheme}b) is one particular example of such graphs, some
other particular examples are shown in Figure \ref{fig:examples}
. We show here that the fixation probability of these graphs is a
simple rational function of the fitness $r$ and of the number of
nodes $M_{A}$ and $M_{B}$ in each class 
\[
\pi_{f}=f(M_{A},M_{B},r)
\]
The exact expression of this function is given by equation (\ref{eq:bithermalfix})
and its plot by figure \ref{fig:pifplot}a. When there is the same
number of nodes in each class ($M_{A}=M_{B}$), bithermal graphs become
isothermal and the function $f$ above is equal to the Moran expression
(\ref{eq:Moran0}). When the imbalance between the number of nodes
in each class is large ($M_{A}\gg M_{B}$ or $M_{A}\ll M_{B}$), the
fixation probability tends toward $0$ or $1-1/r^{2}$, depending
on the nature of the Moran process (birth-death or death-birth).

This article is organized as follow. In the next section, we recall
the continuous time stochastic process of Moran on graph and its associated
Master equation ; the third section is devoted to the Probability
Generating Function method ; In the fourth section, we apply these
results to the bithermal graphs and give their exact fixation probability.
The last section is devoted to some generalizations and conclusions.

\section{Continuous time Moran process on graph.}

Consider a community of $M$ individuals, which can either be wild
type (WT) with fitness $1$ or mutant with relative fitness $r=1+s$.
The individuals are spread spatially and the progeny of an individual
$i$ can replace individual $j$ according to a connectivity map.
The connectivity map can be envisioned as a graph, where each node
$i$ contains exactly one individual, either mutant or wild type ;
the weight of a link $m_{ij}$ specifies the probability for the progeny
of an individual at node $i$ to replace an individual at node $j$
(Figure \ref{fig:EGscheme}a). The coefficients $m_{ij}$ are collected
into a connectivity matrix $\mathbf{m}$. As the number of individual
is fixed, it is sufficient to specify the number of mutants (0 or
1) on each node $\mathbf{n}=(n_{1},n_{2},...n_{M})$ at a given time
to have  complete information about the system at this time. We consider
here a continuous time model where birth (or death) events occur randomly
with rate $\mu$ (\citet{Houchmandzadeh2010}). The probability density
for a node $i$ to decrease or increase its number of mutants by one
unit during a time interval $dt$ is (\citet{Houchmandzadeh2011})
\begin{eqnarray}
W_{i}^{-}(n_{i}) & = & (\mu/M)n_{i}\sum_{k=1}^{M}m_{ki}(1-n_{k})\label{eq:transminus}\\
W_{i}^{+}(n_{i}) & = & (\mu/M)r\left(1-n_{i}\right)\sum_{k=1}^{M}m_{ki}n_{k}\label{eq:transplus}
\end{eqnarray}
It is crucial at this step to distinguish between two kinds of Moran
processes (\citet{Antal2006}). In the first case which we call D-B
(Death first and then replacement, also called Voter Model), a death
occurs first on a node, then immediately one connected node duplicates
and send its progeny to this node. Equation (\ref{eq:transminus})
is therefore the probability density that a mutant dies at node $i$
during $dt$, and is replaced by the progeny of a WT on a connected
node $k$. Equation (\ref{eq:transplus}) is the probability density
that a WT dies at node $i$ and is replaced by a mutant on a connected
node. Without loss of generality, the mutant's advantage $r$ is included
in this line, either as a decreased mortality or a better replacement
success once a death event has occurred. 

In the case of B-D processes (also called Invasion Process), a birth
occurs first on a node $k$ and the progeny is then sent to a connected
node $i$ to replace the local resident. The transition probabilities
are still expressed by the same equations (\ref{eq:transminus},\ref{eq:transplus}),
but the quantity $\mu$ now denotes the birth rate. 

Although the rate equations (\ref{eq:transminus},\ref{eq:transplus})
are similar for these two processes, the normalization conditions
of $m_{ki}$ coefficients are different : 
\begin{equation}
\sum_{i=1}^{M}m_{ik}=1\,\,(\mbox{D-B});\,\,\sum_{i=1}^{M}m_{ki}=1\,\,(\mbox{B-D})\,\,\forall k\label{eq:normalization}
\end{equation}
The temperature of a node is defined for these processes as 
\begin{equation}
\sum_{i=1}^{M}m_{ki}=T_{k}\,\,\mbox{(D-B)\,\,;\,\,}\sum_{i=1}^{M}m_{ik}=T_{k}\,\,(\mbox{B-D})\,\,\forall k\label{eq:temperature}
\end{equation}
Because of the normalization constraint (\ref{eq:normalization}),
we must have $\sum_{j=1}^{M}T_{j}=M$ for both processes. This means
that if some nodes are cold ($T<1$), others must be hot ($T>1$). 

The Moran process is a one-step stochastic one, where during an infinitesimal
interval $dt$, only one birth or death event can occur. Equations
(\ref{eq:transminus},\ref{eq:transplus}) are transition probabilities
between states $\mathbf{n}$ on the one hand and states $a_{i}\mathbf{n}=(n_{1},...n_{i}-1,...n_{M})$
and $a_{i}^{\dagger}\mathbf{n}=(n_{1},...n_{i}+1,...n_{M})$ on the
other. The probability $P(\mathbf{n},t)$ of observing state $\mathbf{n}$
at time $t$ obeys the Master equation 
\begin{equation}
\frac{\partial P(\mathbf{n},t)}{\partial t}=\sum_{\{\mathbf{m}\}}W(\mathbf{m}\rightarrow\mathbf{n})P(\mathbf{m},t)-W(\mathbf{n}\rightarrow\mathbf{m})P(\mathbf{n},t)\label{eq:MasterEquation}
\end{equation}

The EG process we are considering has two absorbing states $\mathbf{1}=(1,1,...1)$
and $\mathbf{0}=(0,0,...0)$. Once a mutant has invaded all the nodes
or has been eliminated from all of them, it is fixed or lost and there
is no further evolution (until a new mutant appears by random mutation)
: $W(\mathbf{1}\rightarrow\mathbf{m})=0$ and $W(\mathbf{0}\rightarrow\mathbf{m})=0$
as  can be deduced from equations (\ref{eq:transminus},\ref{eq:transplus}).
Note that the probability of reaching state $\mathbf{1}$ from an
initial state $\mathbf{n}$, \emph{i.e.} the fixation probability
$\pi(\mathbf{n})$ can in principle be found using the Kolmogorov's
backward equation (\citet{Ewens2004}) : 
\begin{eqnarray}
\sum_{\mathbf{m}}\left(\pi(\mathbf{n})-\pi(\mathbf{m})\right)W(\mathbf{n} & \rightarrow & \mathbf{m})=0\label{eq:fix2}\\
\pi(\mathbf{0})=0\,\,;\,\,\pi(\mathbf{1})=1
\end{eqnarray}
which is a set of linear equations in the unknowns $\pi(\mathbf{n})$.
The direct resolution of the above set of equation however can be
attempted only in special cases. The best example of a direct solution
is the unstructured population, where the graph is fully connected
and $m_{ij}=1/M$ ; the EG dynamics can then be mapped into a biased
one-dimensional Brownian motion and solved by standard techniques
(\citet{Ewens2004}), which yields the well known result (\ref{eq:Moran0})
. Other cases, such as the star topology in the limit of large population
considered by Lieberman et al (\citet{Lieberman2005}), use such careful
mapping. The mapping method however is hard to generalize.

\section{The fp-PGF method. }

Computation of the fixation probability can be simplified if instead
of the linear system (\ref{eq:fix2}), we use the dynamics of the
Probability Generating Function
\[
\phi(\mathbf{z},t)=\sum_{\{\mathbf{n}\}}P(\mathbf{n},t)\mathbf{z}^{\mathbf{n}}
\]
where the variable $\mathbf{z}=(z_{1},z_{2},...z_{M})$ is conjugate
to $\mathbf{n}=(n_{1},n_{2},...,n_{M})$ and $\mathbf{z}^{\mathbf{n}}=z_{1}^{n_{1}}z_{2}^{n_{2}}...z_{M}^{n_{M}}$.
Time is measured in generation time units $M/\mu$. Note that $\phi(\mathbf{0},t)=0$
and $\phi(\mathbf{1},t)=1$. From the Master Equation (\ref{eq:MasterEquation}),
we can derive the dynamics of the PGF (\citet{Houchmandzadeh2011})
which reads: 
\begin{equation}
\frac{\partial\phi}{\partial t}=\sum_{k=1}^{M}f_{k}(\mathbf{z})\frac{\partial\phi}{\partial z_{k}}-\sum_{i,k=1}^{M}g_{i,k}(\mathbf{z})\frac{\partial^{2}\phi}{\partial z_{i}\partial z_{k}}\label{eq:PGF}
\end{equation}
where for D-B processes, 
\begin{eqnarray}
f_{k}(\mathbf{z}) & = & z_{k}\left(\sum_{i=1}^{M}m_{ki}(z_{i}-1)\right)-r^{-1}(z_{k}-1)\label{eq:pgf1-db}\\
g_{i,k}(\mathbf{z}) & = & m_{ki}(z_{i}-1)(z_{i}-r^{-1})z_{k}\label{eq:pgf2-db}
\end{eqnarray}
For the B-D process, the first order term is slightly different and
reads
\[
f_{k}(\mathbf{z})=z_{k}\left(\sum_{i=1}^{M}m_{ki}(z_{i}-1)\right)-(T_{k}/r)(z_{k}-1)
\]
Solving the PGF equation (\ref{eq:PGF}) would seem at least as formidable
as solving directly the Master Equation (\ref{eq:MasterEquation}).
However, for the computation of the fixation probabilities, we are
only interested in the large time limit $t\rightarrow\infty$. At
large time, the mutant is either fixed or lost, therefore the stationary
solution $\phi_{s}(\mathbf{z})$ to which the PGF converges is simply
\begin{equation}
\phi_{s}(\mathbf{z})=\pi_{0}+\pi_{f}\prod_{i=1}^{M}z_{i}\label{eq:PGFstationary}
\end{equation}
where $\pi_{0}$ and $\pi_{f}$ are the loss and fixation probabilities
and implicit functions of the initial conditions. It can also be checked
manually that (\ref{eq:PGFstationary}) is indeed a solution of (\ref{eq:PGF}).
The problem of finding $\pi_{0}$ and $\pi_{f}$ becomes trivial if
the PGF possesses a fixed point $\mathbf{\boldsymbol{\zeta}}=(\zeta_{1},...\zeta_{M})$
such that 
\[
\left.\frac{\partial\phi}{\partial t}\right|_{\mathbf{z}=\boldsymbol{\zeta}}=0
\]
In this case, we have 
\[
\phi(\boldsymbol{\zeta},0)=\phi(\boldsymbol{\zeta},\infty)=\pi_{0}+\pi_{f}\prod_{i=1}^{M}\zeta_{i}
\]
and, as $\pi_{f}+\pi_{0}=1$, 
\[
\pi_{f}=\frac{1-\phi(\boldsymbol{\zeta},0)}{1-\prod_{i=1}^{M}\zeta_{i}}
\]
As the quantity $\phi(\boldsymbol{\zeta},0)$ is known from the initial
conditions, finding a fixed point of the PGF determines entirely the
fixation probability. Note that once a fixed point has been found,
the fixation probability for any initial condition can be trivially
computed. 

For the initial condition of the mutant appearing at random with probability
$1/M$ on one node, $\phi(z_{1},...z_{M};t=0)=(1/M)\sum_{i}z_{i}$
and therefore the fixation probability is given by

\begin{equation}
\pi_{f}=\frac{1-(1/M)\sum_{i}\zeta_{i}}{1-\prod_{i}\zeta_{i}}\label{eq:fixonemutantatrandom}
\end{equation}
The condition for a point $\boldsymbol{\zeta}$ to be a fixed point
is 
\begin{eqnarray}
f_{k}(\boldsymbol{\zeta}) & = & 0\,\,\forall k\label{eq:fixcond1}\\
g_{i,k}(\boldsymbol{\zeta})+g_{k,i}(\boldsymbol{\zeta}) & = & 0\,\,\forall i,k\label{eq:fixcond2}
\end{eqnarray}

Whether such a fixed point exists or not depends on the connectivity
matrix $m_{ik}$. For an isothermal graph where all the nodes $k$
have temperature $T_{k}=1$, it is easy to check that $\boldsymbol{\zeta}=r^{-1}\mathbf{1}$
is a fixed point : 
\begin{eqnarray*}
f_{k}(r^{-1}\mathbf{1}) & = & r^{-1}(r^{-1}-1)-r^{-1}(r^{-1}-1)=0\\
g_{i,k}(r^{-1}\mathbf{1}) & = & m_{ik}(r^{-1}-1)(r^{-1}-r^{-1})r^{-1}=0
\end{eqnarray*}
and the fixation probability (\ref{eq:fixonemutantatrandom}) of the
isothermal graph is equal to the result (\ref{eq:Moran0}) for unstructured
populations. We see here how easily this theorem can be obtained from
the fixed point of the PGF.

\section{bithermal graphs.}

\begin{figure}
\begin{centering}
\includegraphics[width=0.5\columnwidth]{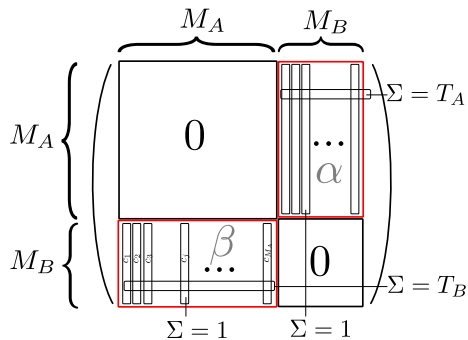}
\par\end{centering}

\caption{The connectivity matrix of a 2-thermal graph for a D-B process, where
nodes of type $A$ are numbered from 1 to $M_{A}$ and nodes of type
$B$ from $M_{A}+1$ to $M_{A}+M_{B}$. For a B-D process, the sum
of the elements of a row is equal to unity, and the sum of the elements
of a column is equal to $T_{A}$ or $T_{B}$. \label{fig:connectmatrix}}

\end{figure}

We now consider a subset of bipartite graphs that  we call bithermal
(\ref{fig:examples}) for which the fixation probability can be determined
in algebraic closed form. In these graphs, there are $M_{A}$ nodes
of type $A$ at temperature $T_{A}$ and $M_{B}$ nodes of type $B$
at temperature $T_{B}$. More over, we require that two nodes can
be connected only if they are at different temperatures, and if a
node $i$ is connected to a node $j$, then $j$ is also connected
to $i$. Of course, we suppose that the graph is connected, \emph{i.e.
}there is always a path from any node $i$ to any node $j$. With
appropriate numbering of the nodes, the connectivity of such graph
is a block matrix of the form
\[
\mathbf{m}=\left(\begin{array}{cc}
0 & \alpha\\
\beta & 0
\end{array}\right)
\]
 (figure \ref{fig:connectmatrix}). The star topology (figure \ref{fig:EGscheme}b)
is such a graph with $M_{A}=1$ central and $M_{B}$ peripheral nodes.
For a D-B process, all the elements of the star's $\beta$ block are
equal to $1/M_{B}$ 
 and all the elements of $\alpha$ block are equal to $1$. Therefore
$T_{A}=M_{B}$ and $T_{B}=1/M_{B}$. The star graph can be generalized
to the case where the number of central nodes $M_{A}>1$ (figure \ref{fig:examples}a).
For a general bithermal graph the number of $B$ nodes to which an
$A$ node connects and the \emph{weights} of these links can be arbitrary,
as long as the constraints on temperatures are respected. Note that
$T_{A}$ and $T_{B}$ are not independent. By summing over all the
elements $m_{ki}$ of the $\beta$ or $\alpha$ block of the connectivity
matrix $\mathbf{m}$, we have for a D-B process 
\[
M_{A}=M_{B}T_{B}\,\,;\,\, M_{B}=M_{A}T_{A}
\]
for a B-D process, the role of $M_{A}$ and $M_{B}$ are exchanged,
but in both cases, we have 
\begin{equation}
T_{A}T_{B}=1\label{eq:tempconst}
\end{equation}

We now search for the bithermal graphs which have an exact fixed point.
Following the example of the isothermal graphs, we look for a fixed
point $\boldsymbol{\zeta=}(\zeta_{1},...\zeta_{M})$ where $\zeta_{i}=\zeta_{A}$
if $i\in A$ and $\zeta_{i}=\zeta_{B}$ if $i\in B$ . 

Let us first consider the case of D-B processes. Using  condition
(\ref{eq:fixcond1}) for $k\in A\,\mbox{or}\, B$ we obtain a set
of two algebraic equations : 
\begin{eqnarray*}
f_{A}(\boldsymbol{\zeta}) & = & \zeta_{A}(\zeta_{B}-1)T_{A}-r^{-1}(\zeta_{A}-1)=0\\
f_{B}(\boldsymbol{\zeta}) & = & \zeta_{B}(\zeta_{A}-1)T_{B}-r^{-1}(\zeta_{B}-1)=0
\end{eqnarray*}
the solution of which is given by 
\begin{eqnarray}
\zeta_{A} & = & u(r,M_{A}/M_{B})\label{eq:DBzA}\\
\zeta_{B} & = & u(r,M_{B}/M_{A})\label{eq:DBzB}\\
u(r,T) & = & \frac{1/r+T}{r+T}\label{eq:defineu}
\end{eqnarray}
This solution also satisfies  condition (\ref{eq:fixcond2}) if for
$k\in A$ and $i\in B$, we have the following relation between the
coefficient of the connectivity matrix $\mathbf{m}$ : 
\begin{eqnarray*}
\frac{m_{ki}}{m_{ik}} & = & -\frac{(\zeta_{B}-1)(\zeta_{B}-r^{-1})\zeta_{A}}{(\zeta_{A}-1)(\zeta_{A}-r^{-1})\zeta_{B}}\\
 & = & \frac{1}{T_{A}}=\frac{M_{A}}{M_{B}}
\end{eqnarray*}
which we can express as a relation between the two blocks $\beta$
and $\alpha$ of the connectivity matrix $\mathbf{m}$: 
\begin{equation}
\alpha=T_{A}\beta^{\intercal}\label{eq:condfix3}
\end{equation}
We observe here that this is the sufficient condition to form exactly
solvable bithermal graphs : form an $M_{B}\times M_{A}$ matrix $\beta$
where the sum of elements in each column is 1 and the sum of elements
in each row is $1/T_{A}=M_{A}/M_{B}$; form the block matrix $\mathbf{m}$
from $\beta$ and $\alpha=T_{A}\beta^{\intercal}$. The $\beta$ block
contains $M_{A}M_{B}$ coefficients subject to $M_{A}+M_{B}$ summation
rules, so the number of bithermal graphs with exact solutions is indeed
very large when $M_{A}$ or $M_{B}$ are large. 

An important subset of bithermal graphs that always has an exact fixed
point is a set we call symmetric bithermal graphs. For these graphs,
all the \emph{existing} links from a node $A$ to a node $B$ (respect.
$B$ to $A$ ) have the same weight $m_{AB}$ (resp. $m_{BA}$). The
generalized star graph (fig \ref{fig:examples}a) belongs to this
subset. For these sets, the weight of a link is determined only by
the number of connected nodes, and each $A$ (resp. $B$) node has
always the same number of neighbors. The symmetric subset automatically
satisfies all the constraints and always has exact fixed points.

Once the fixed point is known, the fixation probability is easily
determined from equation (\ref{eq:fixonemutantatrandom}):
\begin{equation}
\pi_{f}=\frac{1-(M_{A}\zeta_{A}+M_{B}\zeta_{B})/(M_{A}+M_{B})}{1-\zeta_{A}^{M_{A}}\zeta_{B}^{M_{B}}}\label{eq:bithermalfix}
\end{equation}

\begin{figure}
\begin{centering}
\includegraphics[width=0.8\columnwidth]{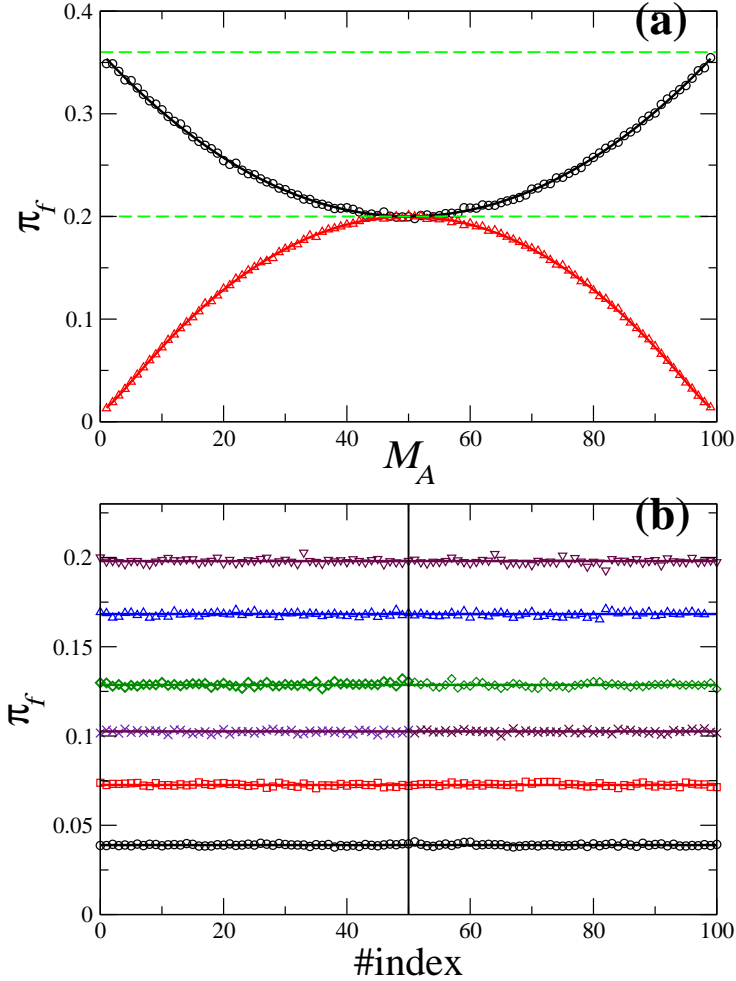}
\par\end{centering}

\caption{Fixation probability $\pi_{f}$ of bithermal EGs of size $M=100$
and $r=1.25$. (a) $\pi_{f}$ as a function of the number $M_{A}$
of nodes of type $A$ for a generalized ($M_{A},M_{B})$ star. circles
(black) and triangles (red) correspond to numerical resolution of
the fixation probabilities for B-D and D-B processes. Solid lines
correspond to exact solutions given by equation (\ref{eq:bithermalfix}).
The dashed lines correspond to $\pi_{f}=1-1/r$ and $\pi_{f}=1-1/r^{2}$.
Numerical simulations were performed by using a Gillespie Algorithm(\citet{Gillespie1977})
; for each point, $10^{5}$ stochastic paths were generated and used
to compute the fixation probability. (b)Fixation probability $\pi_{f}$
for random bithermal EG for different values of $M_{A}$ (circles
5, squares 10, $\times$ 15, diamond 20, triangle up 30,triangle down
45) and D-B processes. To the left of the vertical line (\#index $\le$50)
: Each point corresponds to a random bithermal connectivity matrix
where the two blocks are related through $\alpha=T_{A}\beta^{\intercal}$.
To the right of the vertical line (\#index >50) : For each point,
the two blocks of the bithermal connectivity matrix are random and
$\alpha\ne T_{A}\beta^{\intercal}$. Solid lines indicate the theoretical
values computed by equation (\ref{eq:bithermalfix}).\label{fig:pifplot}}

\end{figure}

For a B-D process, the computation of the fixed point follows exactly
the same line of argument. The result is obtained by permuting $M_{A}$
and $M_{B}$ : 
\begin{eqnarray}
\zeta_{A}^{\mbox{BD}} & = & u(r,M_{B}/M_{A})\,\,;\,\,\zeta_{B}^{\mbox{BD}}=u(r,M_{A}/M_{B})\label{eq:BDzAB}
\end{eqnarray}
The fixation probability is given by the same expression (\ref{eq:bithermalfix}).
Note that D-B and B-D processes act in opposite directions. Compared
to a non-structured population of the same size, a bithermal D-B process
acts as a suppressor of selection, lowering the fixation probability
where the B-D process is an amplifier of selection, increasing the
fixation probability. The maximum $\pi_{f}$ for a D-B is obtained
for $T_{A}=1$ and is equal to expression (\ref{eq:Moran0}) ; the
maximum $\pi_{f}$ for a B-D process is obtained for $T_{A}\rightarrow\infty$
and is equal to $1-r^{-2}$. A plot of both fixation probabilities
and their comparison to numerical simulations is shown in figure \ref{fig:pifplot}
. 

We stress that for bithermal graphs, the details of the connectivity
are not important : the fixation probability depends only on the total
number of $A$ and $B$ nodes. Consider for example the symmetric
generalized star $(M_{A},M_{B},p,q)$ where each $A$ node is connected
to $p$ nodes of type $B$ and each $B$ nodes to $q$ nodes of type
$A$. For a fully connected generalized star, $p=M_{B}$ and $q=M_{A}$
; an example of partially connected generalized star is given in figure
\ref{fig:examples}a. We emphasize that for generalized stars, the
fixation probability does not depend on the detail of the connectivity
$p,q$, a result which would be hard to predict by other methods.
We also note from numerical simulations that the \emph{fixation time}
is also only a function of $M_{A}$ and $M_{B}$ and does not depend
on the detail of the connectivity. 

Finally, we note that even when the constraint (\ref{eq:condfix3})
is not respected and $\alpha\ne T_{A}\beta^{\intercal}$, the fixation
probability of bithermal graphs is well approximated by expression
(\ref{eq:bithermalfix}). In this case, the point $\boldsymbol{\zeta}$
computed from equations (\ref{eq:DBzA},\ref{eq:DBzB}) is only a
quasi-fixed point, but as we have shown earlier(\citet{Houchmandzadeh2011}),
for large communities, quasi-fixed points can be used for a good approximation
of the fixation probability. It can be observed in figure (\ref{fig:pifplot}b)
that the numerical errors of fixation probabilities of connectivity
matrices having exact fixed points (left of vertical line) or only
quasi-fixed point (right of vertical line) are of the same magnitude,
for a system as small as $M=100$. We also note from numerical simulations
that \emph{fixation time}, is only

\section{Discussion and Conclusion.}

\begin{figure}
\begin{centering}
\includegraphics[width=0.85\columnwidth]{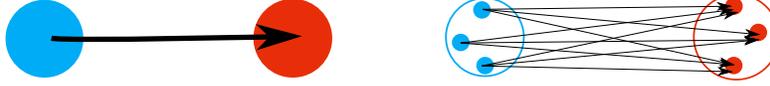}
\par\end{centering}

\caption{Generalization to bi-level graphs, where each node can be considered
as composed of $N$ individuals, conserving the same topology, \emph{i.e.}
the weight of new links is downscaled by a factor $N$ : $m'_{ij}=m_{ij}/N$
\label{fig:bilevel}}

\end{figure}

The approach we presented above can be generalized in various directions.
We have restricted our approach to the case where each node contains
only one individual. This restriction can be relaxed and we can let
each node contain a number of individuals $N\ge1$ (figure \ref{fig:bilevel})
. This is equivalent to the original island model of Maruyama(\citet{Maruyama1974})
or can alternatively be envisioned as a bi-level graph as defined
by Shakarian \emph{et al}(\citet{Shakarian2012}), where each node
is mapped into $N$ sub-nodes conserving the original topology. As
we have shown earlier(\citet{Houchmandzadeh2011}), this parameter
does not alter the expression of the PGF and the fixed points are
still computed by the same equations. The fixation probability is
slightly modified and reads 
\[
\pi_{f}=\frac{1-(M_{A}\zeta_{A}+M_{B}\zeta_{B})/(M_{A}+M_{B})}{1-\zeta_{A}^{NM_{A}}\zeta_{B}^{NM_{B}}}
\]
The approach can be extended even further and allows for different
numbers $N_{k}$ of individuals for each node $k$ (\citet{Houchmandzadeh2011}).

Another direction toward which this work can be extended is the $n-$thermal
graphs. Here we have considered only bithermal graphs composed of
two types of nodes which we can represent by an $A-B$ topology. In
principle, we can generalize the method to graphs containing $P$
\emph{types} of nodes $O_{1},...,O_{P}$ : The $M_{I}$ nodes belonging
to type $O_{I}$ have the temperature $T_{I}$ and a hot class can
only be connected to cold classes and vice versa. We could in principle
form a polymeric topology such as $O_{1}-O_{2}...-O_{P}$, branched
systems, closed rings and so on. The exploration of these topologies
implies an analytic study of the roots of algebraic equations of order
$2P$ and is  beyond the scope of the present article.

To summarize, we have obtained exact analytical results for a wide
class of graphs that we have called bithermal in the field of Evolutionary
Graph Theory. EGT is a cornerstone for our understanding of evolution,
because natural population are always geographically extended and
cannot be a priori approximated as ``well mixed''. Exact results
in EGT however have been hard to obtain because in each case, a mapping
into a one-dimensional Brownian motion has to be constructed ; whether
such a mapping exists or not for a particular problem is not a trivial
problem. The method we develop is radically different : by using the
continuous time version of the Moran model and the dynamics of the
PGF, we reduce the problem of finding an exactly solvable model into
finding the roots of algebraic equations. We have illustrated the
power of this dynamical method through our study of bithermal graphs.
We believe that the method we have presented can be a powerful tool
to get exact results for the fixation probability of more complex
evolutionary graphs.

\subsection*{Acknowledgements.}

We are grateful to E. Geissler and O. Rivoire for the careful reading
of the manuscript and fruitful discussions. This work was partly funded
by Agence Nationale de la Recherche Française (ANR) grant \textquotedblleft{}Evo-Div.\textquotedblright{}

\bibliographystyle{elsarticle-harv}

\end{document}